\documentclass[12pt]{article}   
\usepackage{epsfig}
\newcommand{\mysection}{\setcounter{equation}{0}\section}

\def\beq{\begin{equation}}
\def\eeq{\end{equation}}
\def\beqa{\begin{eqnarray}}
\def\eeqa{\end{eqnarray}}

\newlength{\dinwidth} \newlength{\dinmargin}
\begin{document}

\begin{center}
{\Large \bf $W$-boson production with large transverse momentum at the LHC}
\end{center}
\vspace{2mm}
\begin{center}
{\large Nikolaos Kidonakis\footnote{Presented at the  XIV International 
Workshop on Deep Inelastic Scattering (DIS 2006), Tsukuba, Japan, 
April 20-24, 2006.}}\\
\vspace{2mm}
{\it Kennesaw State University, Physics \#1202 \\
1000 Chastain Rd., Kennesaw, GA 30144-5591, USA}
\end{center}
\vspace{3mm}
\begin{center}
{\large Richard J. Gonsalves}\\
\vspace{2mm}
{\it Department of Physics, University at Buffalo\\
The State University of New York, 
Buffalo, NY 14260-1500, USA}
\end{center}
\vspace{3mm}
\begin{center}
{\large Agustin Sabio Vera}\\
\vspace{2mm}
{\it Physics Department, Theory Division, CERN\\ 
CH-1211 Geneva 23, Switzerland}  
\end{center}

\vspace{4mm}

\begin{abstract}
We study $W$-boson production with large transverse momentum, 
$Q_T$, in $pp$ collisions at the LHC.
We calculate the complete NLO corrections and the soft-gluon NNLO 
corrections to the differential cross section. 
The NLO corrections are large but they do not reduce the scale dependence
relative to LO, while the NNLO soft-gluon corrections, 
although small, significantly reduce the scale dependence
and thus provide a more stable result.
\end{abstract}

\thispagestyle{empty} \newpage \setcounter{page}{2}

\mysection{Introduction}

$W$ hadroproduction at large $Q_T$ is useful in testing the Standard Model 
and in estimating backgrounds to Higgs production and to new physics
such as new gauge bosons. 
Accurate theoretical predictions for 
$W$ production at the LHC, scheduled to begin operation in 2007,
are thus needed.

Calculations of the NLO
cross section for $W$ production at large
transverse momentum at the Fermilab Tevatron collider were
presented in Refs. [1,2].
The NLO corrections contribute to enhance the
differential distributions in $Q_T$ of the $W$ boson
and they reduce the factorization and renormalization
scale dependence of the cross section at the Tevatron.
More recent studies \cite{NKVD,NKASV} included soft-gluon corrections
through NNLO, 
which provide additional enhancements and a further reduction
of the scale dependence \cite{NKASV}.

Here we discuss $W$ production with large $Q_T$ at the LHC.  
The results presented are based on Ref.~[5].
The partonic channels at LO are 
$q(p_a) + g(p_b) \longrightarrow W(Q) + q(p_c)$
and  
$q(p_a) + {\bar q}(p_b) \longrightarrow W(Q) + g(p_c)$.  
We define the kinematical invariants 
$s=(p_a+p_b)^2$, $t=(p_a-Q)^2$, $u=(p_b-Q)^2$
and $s_2=s+t+u-Q^2$.
At threshold $s_2 \rightarrow 0$.
The NLO cross section is 
\beqa 
E_Q\,\frac{d\hat{\sigma}_{f_af_b{\rightarrow}W(Q)+X}}{d^3Q}&=&
\delta(s_2)\alpha_s(\mu_R^2)\left[A(s,t,u) \right.
\nonumber \\ && \left. \hspace{-15mm}
{}+\alpha_s(\mu_R^2)
B(s,t,u,\mu_R)\right] + \alpha_s^2(\mu_R^2)C(s,t,u,s_2,\mu_F).
\eeqa
$A(s,t,u)$ arises from the LO processes. 
$B(s,t,u,\mu_R)$ is the sum of virtual corrections and of singular terms
${\sim}\delta(s_2)$ in the real radiative corrections.  
$C(s,t,u,s_2,\mu_F)$ is from real emission processes away from $s_2=0$. 
 
The soft-gluon corrections \cite{NKres} are of the form 
$[\ln^l(s_2/Q_T^2)/s_2]_+$, where 
for the order $\alpha_s^n$ corrections $l\le 2n-1$.
These corrections can be calculated at higher orders using the 
formulas in Ref. [7], which have also been applied recently to other 
electroweak processes \cite{ew}. 

\section{Numerical results}

We consider $W$ production at large transverse momentum 
in $pp$ collisions at the LHC with $\sqrt{S}=14$ TeV. 
We use the MRST2002 parton densities \cite{MRST}.

\begin{figure}
\begin{center}
\includegraphics[width=10.0cm]{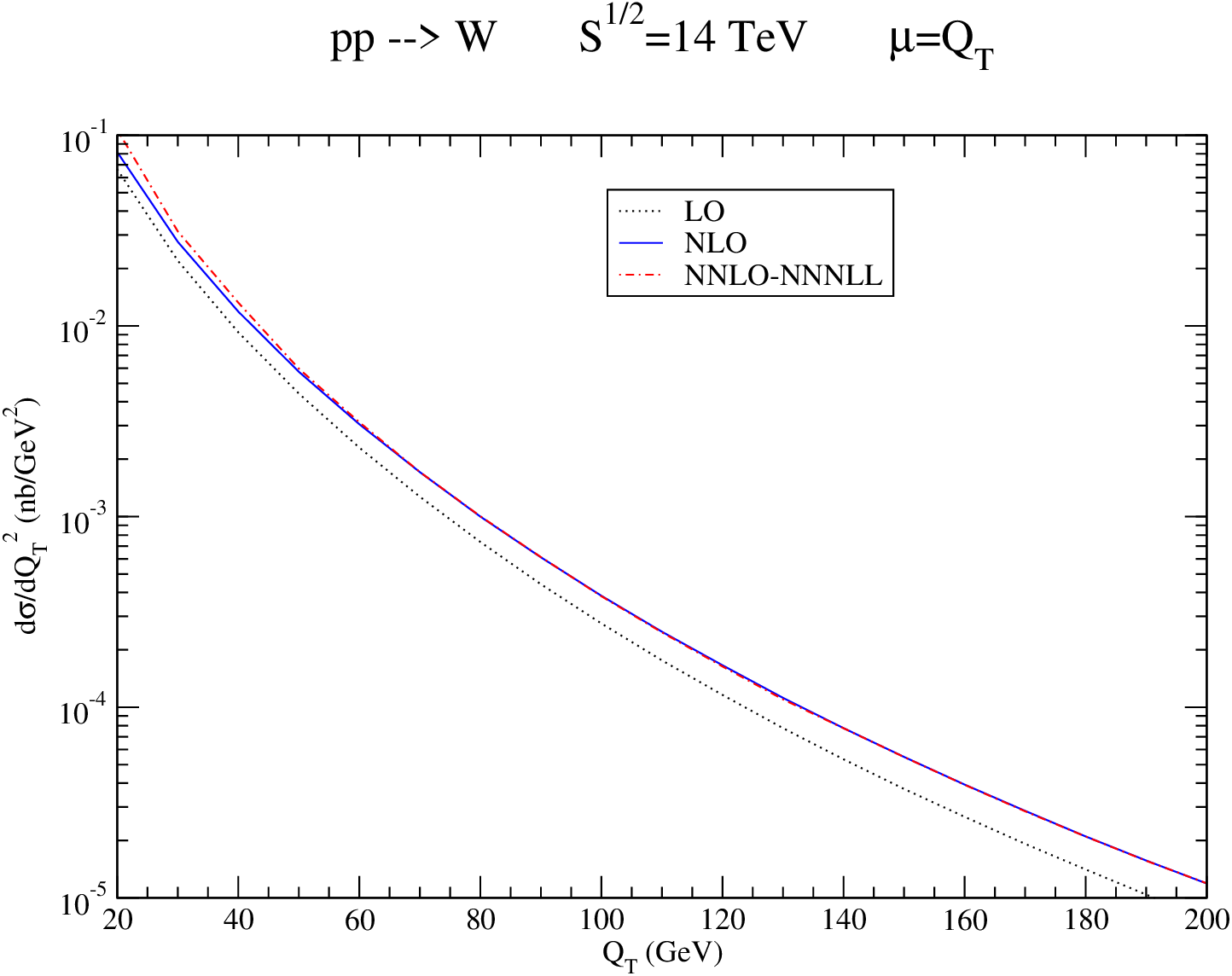}   
\caption{The differential cross section,
$d\sigma/dQ_T^2$, for $W$ production at the LHC.} 
\label{wlhc}
\end{center}
\end{figure}

In Figure 1 we plot the transverse momentum distribution,
$d\sigma/dQ_T^2$, at large $Q_T$. 
Here we set $\mu_F=\mu_R=Q_T$ and denote this common
scale by $\mu$. We plot the LO, NLO, and NNLO-NNNLL results. 
Here NNLO-NNNLL means that 
we include the (approximate) NNNLL soft-gluon terms at NNLO in $\alpha_s$. 
The NLO corrections provide a significant enhancement of the LO $Q_T$
distribution, a 30\% to 50\% increase in the $Q_T$ range shown. 
The NNLO-NNNLL corrections provide a further
rather small enhancement which is hardly visible
in the plot.

\begin{figure}
\begin{center}
\includegraphics[width=10.0cm]{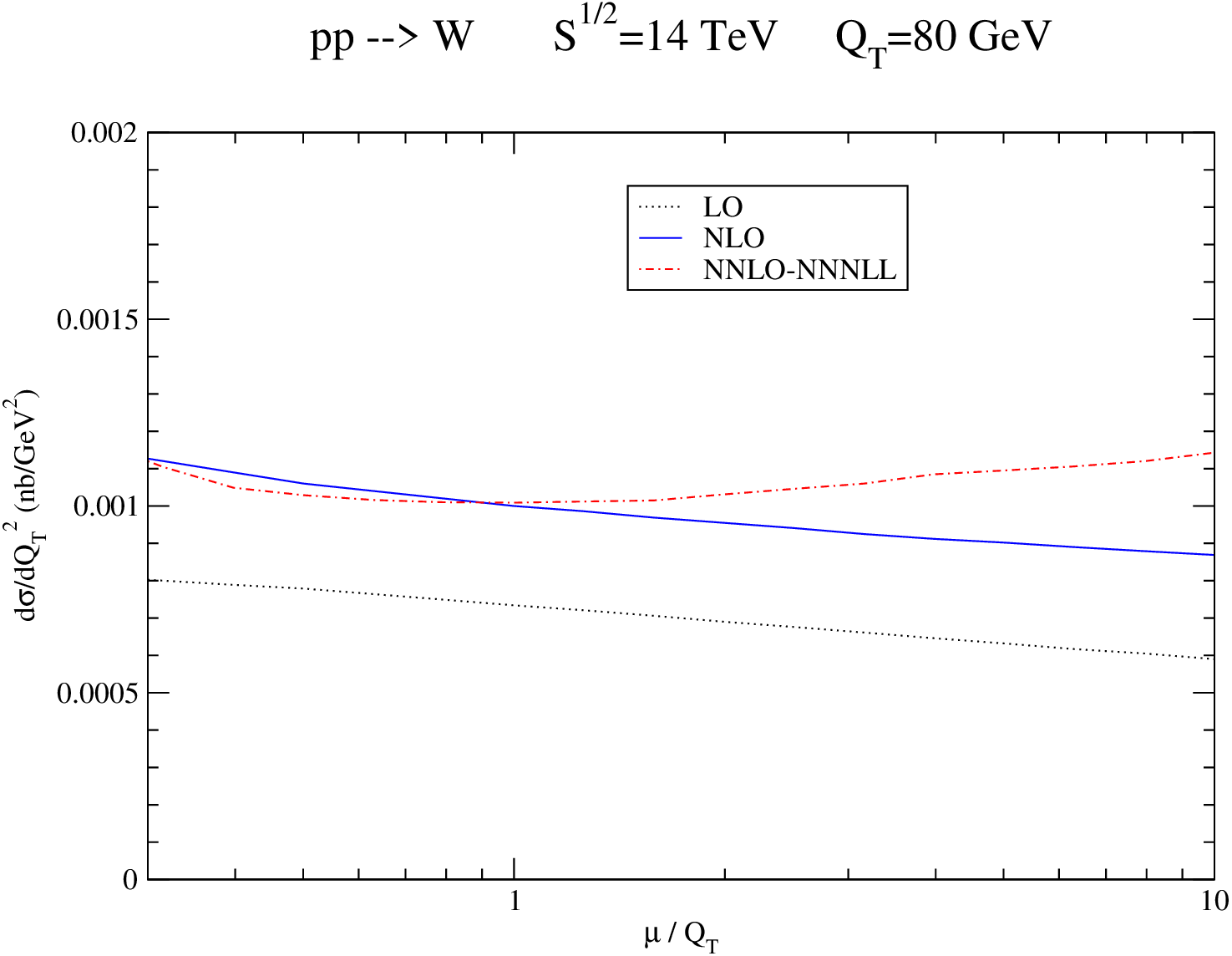}
\caption{$d\sigma/dQ_T^2$ for $W$ production
at the LHC with $Q_T=80$ GeV and $\mu=\mu_F=\mu_R$.} 
\label{mu80wlhc}
\end{center}
\end{figure}

\begin{figure}
\begin{center}
\includegraphics[width=10.0cm]{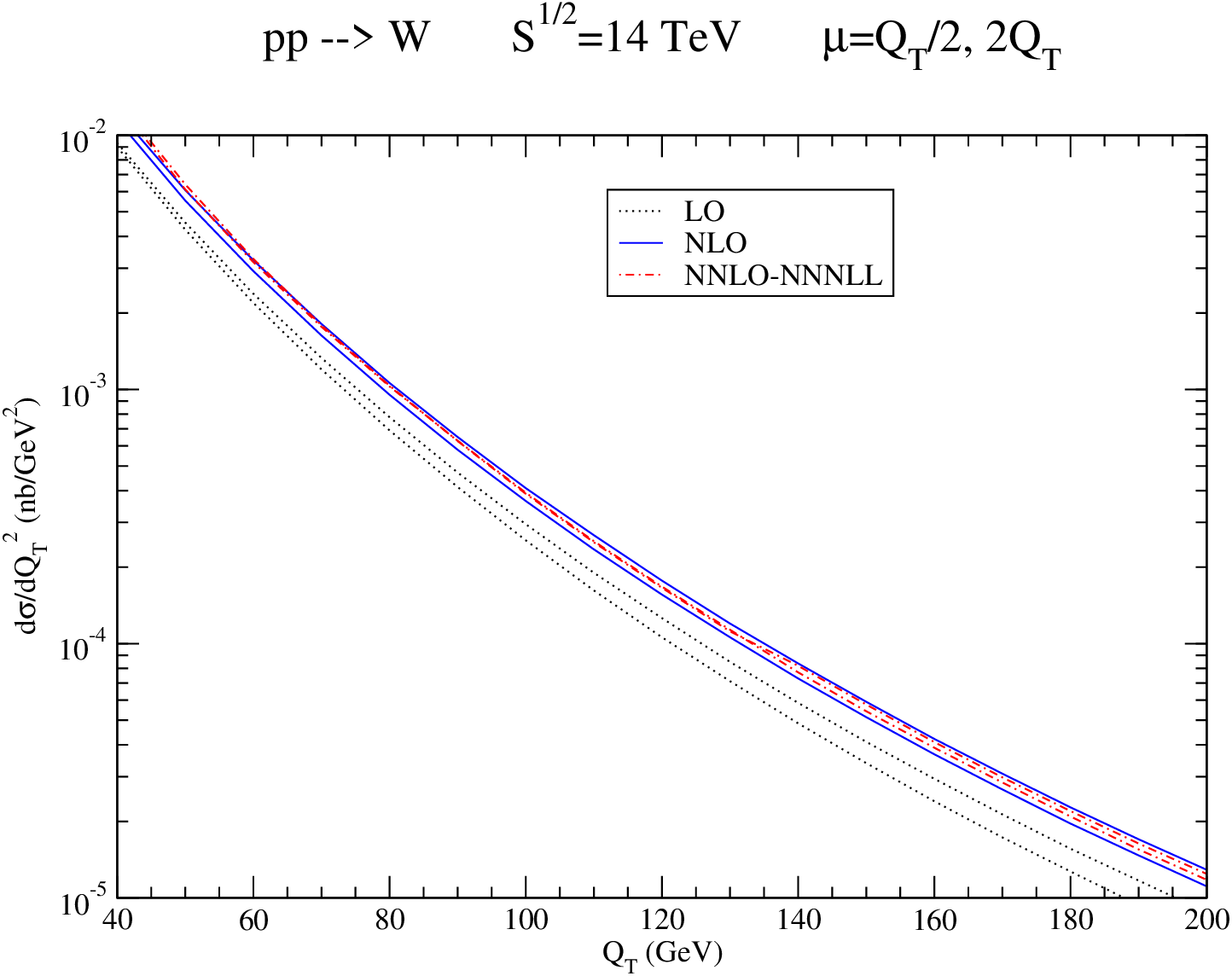}
\caption{$d\sigma/dQ_T^2$ for $W$ production
at the LHC with  $\mu=\mu_F=\mu_R=Q_T/2$ (upper lines) and $2Q_T$
(lower lines).} 
\label{wmuQt}
\end{center}
\end{figure}

In Figure 2
we plot the scale dependence of $d\sigma/dQ_T^2$
for $Q_T=80$ GeV. We note that, surprisingly, the scale dependence
of the cross section is not reduced when the NLO corrections are
included, but we have an improvement when the NNLO-NNNLL corrections
are added. We find similar results for other $Q_T$ values.
If we plot the LO scale dependence separately for
$\mu_F$ and $\mu_R$ with the other held fixed (see fig. 4 of Ref. [5]), 
we find that 
the cross section increases with positive curvature
as the renormalization scale $\mu_R$  is decreased 
(as expected due to asymptotic freedom), but that the 
$\mu_F$ dependence has negative curvature and the cross section
increases with scale.
The latter behavior is due to the fact that the cross section is dominated by
$qg\rightarrow Wq$ and the gluon density in the proton increases rapidly with 
scale at fixed $x$ smaller than ${\sim}0.01$.
At LHC energies, the $\mu_R$ and $\mu_F$ dependencies cancel one another
approximately.

In Figure 3 we plot $d\sigma/dQ_T^2$ at high $Q_T$ 
for two values of the scale, $Q_T/2$ and $2Q_T$,
often used to display the uncertainty due to scale variation.
We note that while the variation of the LO cross section is
significant and the variation at NLO is similar to LO, at
NNLO-NNNLL it is very small: the two NNLO-NNNLL curves
lie very close to each other.


\begin{thebibliography}{0}

\bibitem{AR}
P.B. Arnold and M.H. Reno,  {\it Nucl. Phys.} {\bf B319}, 37 (1989);
(E) {\bf B330}, 284 (1990).
  
\bibitem{gpw}
R.J. Gonsalves, J. Pawlowski, and C.-F. Wai, {\it Phys. Rev.} {\bf D40},
2245 (1989); {\it Phys. Lett.} {\bf B252}, 663 (1990).

\bibitem{NKVD}
N. Kidonakis and V. Del Duca,  {\it Phys. Lett.} {\bf B480}, 87 (2000).

\bibitem{NKASV}
N. Kidonakis and A. Sabio Vera, {\it JHEP} {\bf 02}, 027 (2004).

\bibitem{GKS}
R.J. Gonsalves, N. Kidonakis, and A. Sabio Vera, {\it Phys. Rev. Lett.}
{\bf 95}, 222001 (2005).

\bibitem{NKres}
N. Kidonakis and G. Sterman, {\it Phys. Lett.} {\bf B387}, 867 (1996);
{\it Nucl. Phys.} {\bf B505}, 321 (1997).

\bibitem{NKuni}
N. Kidonakis, {\it Int. J. Mod. Phys.} {\bf A19}, 1793 (2004);
{\it Mod. Phys. Lett.} {\bf A19}, 405 (2004);
{\it Phys. Rev.} {\bf D73}, 034001 (2006).

\bibitem{ew}
N. Kidonakis, {\it JHEP} {\bf 05}, 011 (2005);
N. Kidonakis and A. Belyaev, {\it JHEP} {\bf 12}, 004 (2003).

\bibitem{MRST}
A.D. Martin, R.G. Roberts, W.J. Stirling, and R.S. Thorne,
Eur. Phys. J. C {\bf 28}, 455 (2003).

\end{thebibliography}
\end{document}